\documentclass[journal=jacsat,manuscript=article,layout=twocolumn]{achemso}

\usepackage[version=3]{mhchem} 
\usepackage{amssymb}
\usepackage{amsfonts}
\usepackage{xcolor}

\newcommand{\onlinecite}[1]{[\hspace{-1 ex} \nocite{#1}\citenum{#1}]} 

\author{Gauthier Legrand}
\affiliation{ENSL, CNRS, Laboratoire de Physique, F-69342 Lyon, France.}
\author{Guilhem P. Baeza}
\affiliation{Univ Lyon, INSA Lyon, UCBL, CNRS, MATEIS, UMR5510, 69621, Villeurbanne, France.}
\author{Matteo Peyla}
\affiliation{ENSL, CNRS, Laboratoire de Physique, F-69342 Lyon, France.}
\author{Lionel Porcar}
\affiliation{Institut Laue Langevin, 38000 Grenoble, France.}
\author{Carlos Fernandez-de-Alba}
\affiliation{Université de Lyon, CNRS, Université Claude Bernard Lyon 1, INSA Lyon, Université Jean Monnet, UMR 5223, Ingénierie des Matériaux Polymères, Service RMN Polymères de l’ICL F-69621 Cédex, France.}
\author{Sébastien Manneville}
\affiliation{ENSL, CNRS, Laboratoire de Physique, F-69342 Lyon, France.}
\alsoaffiliation[IUF]
{Institut universitaire de France (IUF).}
\author{Thibaut Divoux}
\affiliation{ENSL, CNRS, Laboratoire de Physique, F-69342 Lyon, France.}
\email{Thibaut.Divoux@ens-lyon.fr}

\title[An \textsf{achemso} demo]
  {Acid-induced gelation of carboxymethylcellulose solutions}

\keywords{carboxymethylcellulose, gel, Relaxometry, SANS}

\begin{document}







\begin{abstract}
The present work offers a comprehensive description of the acid-induced gelation of carboxymethylcellulose (CMC), a water-soluble derivative of cellulose broadly used in numerous applications ranging from food packaging to biomedical engineering. Linear viscoelastic properties measured at various pH and CMC contents allow us to build a sol-gel phase diagram, and show that CMC gels exhibit broad power-law viscoelastic spectra that can be rescaled onto a master curve following a time-composition superposition principle. These results demonstrate the microstructural self-similarity of CMC gels, and inspire a mean-field model based on hydrophobic inter-chain association that accounts for the sol-gel boundary over the entire range of CMC content under study. Neutron scattering experiments further confirm this picture and suggest that CMC gels comprise a fibrous network crosslinked by aggregates. Finally, low-field NMR measurements offer an original signature of acid-induced gelation from the solvent perspective. Altogether, these results open avenues for precise manipulation and control of CMC-based hydrogels. 

\end{abstract}

\section{Main text}

\begin{figure*}[t]
    \includegraphics[width = 0.8\textwidth]{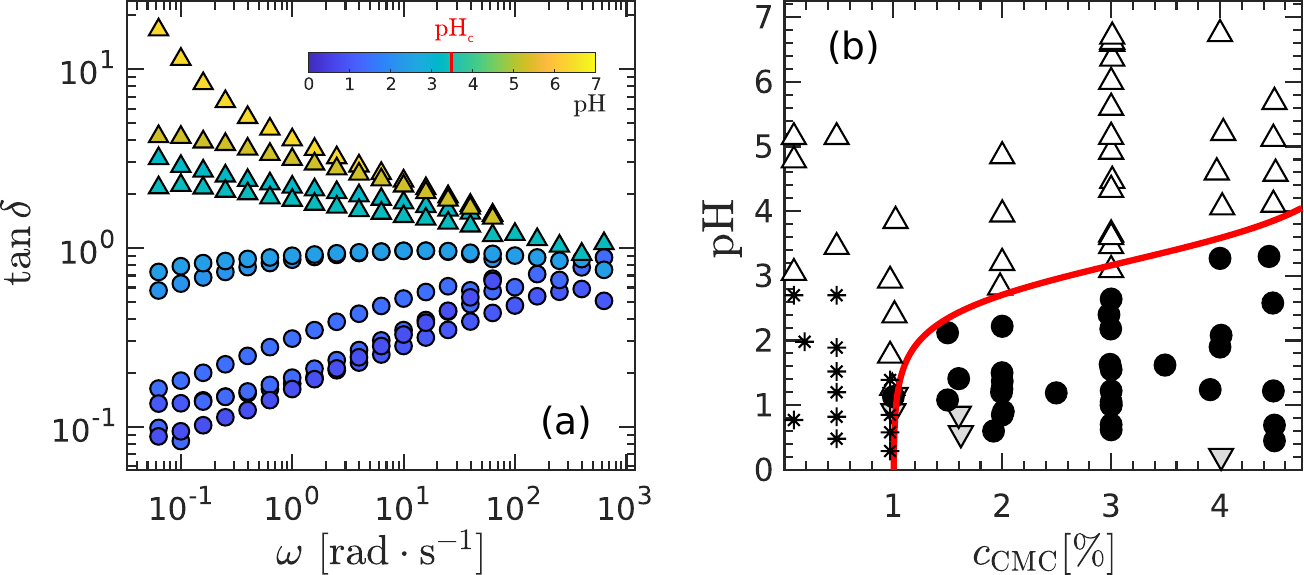}
    \caption{(a) Loss factor $\tan \delta=G''/G'$ vs.~frequency $\omega$ for 3\% CMC solutions at various pH ranging from 0.6 to 6.7 (see color bar). Triangles stand for viscoelastic liquid ($\tan \delta>1$ at low frequencies), whereas circles correspond to gels ($\tan \delta<1$ at low frequencies). (b) Phase diagram pH vs.~CMC concentration comprising {four} phases: {($*$) phase-separated samples}, ($\triangle$) viscoelastic liquids, ($\bullet$) gels, and ($\triangledown$) Newtonian solutions. The latter case corresponds to over-acidified CMC solutions in which the CMC is chemically degraded. The red curve separating the sol and gel phases corresponds to the model discussed briefly in the main text (see full details in the SI).}
    \label{fig:transition_sol_gel}
\end{figure*}

The sodium salt of carboxymethylcellulose (NaCMC) is a water-soluble derivative of cellulose, the most abundant polymer on earth, which is broadly used for industrial applications such as food, pharmaceuticals, paints, etc., and serves as a thickener and water retention agent when dispersed in a solvent.\cite{Clasen:2001,Sheng:2021,Yaradoddi:2020} 
The properties of NaCMC can be tuned via the degree of substitution (DS), which corresponds to the average number of carboxymethyl groups per repeating glucose unit and varies between 0 and 3. Highly substituted polymers, i.e., for DS $\gtrsim 1$, are hydrophilic and disperse easily in water, yielding rheological features typical of polyelectrolyte solutions.\cite{Yang:2007b,Lopez:2017,Behra:2019,Lopez:2020} 
In contrast, weakly substituted polymers, i.e., for DS $\lesssim 0.9$, contain hydrophobic regions, which favor interchain aggregation and the formation of so-called ``fringed micelles'' in aqueous solution,\cite{Liebert:2001} yielding thixotropic and even gel-like properties at high enough concentrations.\cite{Ott:1956,Debutts:1957,Elliot:1974,Hakert:1989,Kastner:1997,Barba:2002}
In practice, the gelation of NaCMC solution can be induced by lowering the pH, which decreases the charge density along the CMC chain and promotes the formation of multichain aggregates.\cite{Dogsa:2014} At a low enough pH, NaCMC solutions thus behave as soft solids that experience a solid-to-liquid transition at large deformations.\cite{Hermans:1965,Lopez:2021}  While much is known about the flow properties of highly substituted NaCMC aqueous solutions, the acid-induced gelation of less substituted CMC solutions, and the resulting gel properties remain poorly understood in terms of mechanical and structural properties.

Here, combining complementary techniques, we perform an in-depth characterization of the acid-induced sol-gel transition in aqueous solutions of weakly substituted NaCMC. Rheological measurements allow us to identify the gelation point and build a phase diagram over a broad range of pH and CMC content. Moreover, the scattered intensity measured by small-angle neutron scattering (SANS) shows an abrupt change at the sol-gel transition, suggesting the presence of a heterogeneous percolated network of disordered fibrils formed through a microphase separation. 
Building upon this scenario, we devise a simple model based on physical crosslinks at the locus of hydrophobic patches along the polymer chain, which yields an equation for the sol-gel boundary in excellent agreement with the phase diagram. 
Viscoelastic spectra measured in the gel phase at different pH exhibit similar features. Following a time-composition superposition principle, these spectra can be rescaled onto a remarkable master curve, which hints at a self-similar microstructure in which the CMC chains follow a Rouse-like dynamics. Finally, low-field NMR spectroscopy reveals that the sol-gel transition is associated with a decrease in the proton transverse relaxation time $T_2$, whose temperature dependence remains Arrhenius-like across the transition. Yet, the corresponding energy scale reveals a sharp drop by more than 60\% at the transition, showing that the aqueous solvent is affected in the \textit{bulk} by the sol-gel transition, echoing the excellent water retention capacity of CMC-based gels.\cite{Xiao:2001,Kono:2014,Marliere:2012}

Aqueous solutions of NaCMC for DS = {0.88 (see SI)} are prepared at ambient temperature, and their linear viscoelastic properties are measured by rheometry following the protocols described in the \textit{Experimental} section. 
Figure~\ref{fig:transition_sol_gel}(a) shows the linear viscoelastic spectrum in the form of the loss factor, i.e., $\tan \delta = G''/G'$ vs.~$\omega$, for a series of 3\%~wt.~CMC solutions with different pH values ranging between pH=0.6 and 6.7. 
For pH~$\geq3$, the loss factor is a decreasing function of the frequency, with $\tan \delta>1$ for vanishing frequencies, i.e., the sample response is mainly liquid-like. In contrast, for pH~$<3$, the loss factor increases for increasing frequency, with $\tan \delta<1$ at low frequencies, which is the hallmark of a solid-like response. These results highlight a critical pH value, pH$_c\simeq 3$, at which the loss factor is independent of the frequency, hence demonstrating that the present NaCMC solution displays a sol-gel transition at pH$_c$.\cite{Winter:1997} Note that for vanishing pH (i.e., $\mathrm{pH}\lesssim 1$), the CMC solutions behave as Newtonian liquids due to the {chemical} degradation of the polymer {(see SI for rheological characterization)}.\cite{Durig:1950,Cheng:1999}

\begin{figure}[!t]
    \includegraphics[width = 0.9\columnwidth]{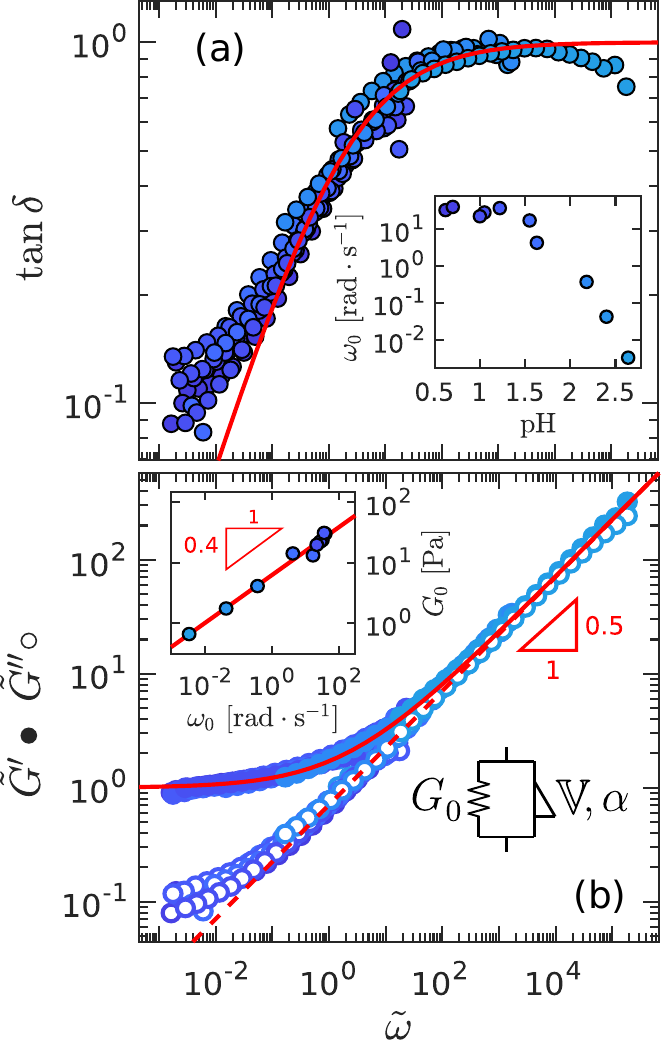}
    \caption{(a) Master curve for the loss factor $\tan \delta$ versus reduced frequency $\tilde{\omega} =\omega/\omega_0$ for a 3\% CMC gels at various $\mathrm{pH}< \mathrm{pH}_c$ ranging between 0.6 and 2.6. Inset: shift factor $\omega_0$ vs pH. The red curve is the best fit of the data by a fractional Kelvin-Voigt  (FKV) model sketched in (b) (lower inset). (b) Master curve for the frequency dependence of the normalized viscoelastic moduli of acid-induced CMC gels. The red curve is the best fit of the data by the FKV model. Upper inset: $G_0$ vs.~$\omega_0$; the red line is the best power-law fit of the data, yielding an exponent 0.4.}
    \label{fig:Rheo_gel}
\end{figure}

Repeating this analysis over a broad range of CMC concentrations allows us to build the phase diagram pH vs.~CMC content reported in Fig.~\ref{fig:transition_sol_gel}(b). The gel-like region is observed beyond a critical CMC content of about 1~\%, and over a growing span of pH for increasing CMC concentration. The equation of the sol-gel phase boundary can be derived using a mean-field approach, in which gelation occurs beyond a critical density of active hydrophobic patches, which serve as cross-linkers. Assuming that the number of active hydrophobic patches increases with both the CMC content and the number of carboxymethyl group, we can derive an equation $\textrm{pH}_{\rm c}(c_{\rm CMC})$ that depends only on three parameters (see SI for full details). The result shown as a red curve in Fig.~\ref{fig:transition_sol_gel}(b) is in excellent agreement with the available data. 

Moreover, in the gel state, the loss factor always displays a power-law regime at low frequency, followed by a high-frequency plateau [see Fig.~\ref{fig:transition_sol_gel}(a)]. This prompts us to fit the viscoelastic spectra for $\mathrm{pH}< \mathrm{pH}_c$ by a compact mechanical model, namely a fractional Kelvin-Voigt (FKV) model,\cite{Legrand:2023} which is sketched as an inset in Fig.~\ref{fig:Rheo_gel}(b). It consists of a spring $G_0$ in parallel with a spring-pot --or Scott Blair-- element \cite{Jaishankar:2013,Bonfanti:2020} characterized by a dimensionless exponent $\alpha$, and a quasi-property $\mathbb{V}$ (with dimension Pa.s$^\alpha$). 
Remarkably, all $\tan \delta(\omega)$ data obtained in the gel phase by varying the pH between 0.5 and 2.5 at fixed $c_{\rm CMC}=3\%$ can be captured by varying $G_0$ and $\mathbb{V}$, while setting $\alpha=1/2$. 
The FKV model is characterized by a single timescale defined as $\omega_0=(G_0/\mathbb{V})^{1/\alpha}$, which allows us to collapse all the loss factor data onto a master curve when plotted against $\tilde \omega =\omega/\omega_0$ [Fig.~\ref{fig:Rheo_gel}(a)]. Moreover, using normalized moduli, i.e., $\tilde G'=G'/G_0$, $\tilde G''=G''/G_0$, we also uncover a universal master curve for the viscoelastic spectrum of all gels obtained at $\mathrm{pH}< \mathrm{pH}_c$ [Fig.~\ref{fig:Rheo_gel}(b)] {(see SI for individual spectra and $G_0$ and $\omega_0$ vs.~pH)}. This master curve is, in turn, well described over 8 orders of magnitude of reduced frequency by the fractional Kelvin-Voigt model pictured as red curves in Fig.~\ref{fig:Rheo_gel}(b). 

These results show that irrespective of the pH, acid-induced CMC gels display a common underlying hierarchical microstructure, which is characterized by the power-law dependence of $G_0$ vs.~$\omega_0$ with an exponent 0.4 [inset in Fig.~\ref{fig:Rheo_gel}(b)] much smaller than the value reported for other polymer gels, e.g., protein gels.\cite{Costanzo:2020} Moreover, in the high-frequency limit, $\tilde G' \sim \tilde G'' \sim \tilde \omega^{1/2}$, which is reminiscent of the power-law rheology of critical gels,\cite{Winter:1997} and in remarkable agreement with a Rouse scaling. This high-frequency response also contrasts with the $2/3$ scaling exponent reported for CMC (and polyacrylamide) hydrogels crosslinked by colloidal particles.\cite{Pashkovski:2003,Legrand:2023,Adibnia:2017}

\begin{figure}[!t]
    \includegraphics[width = 0.9\columnwidth]{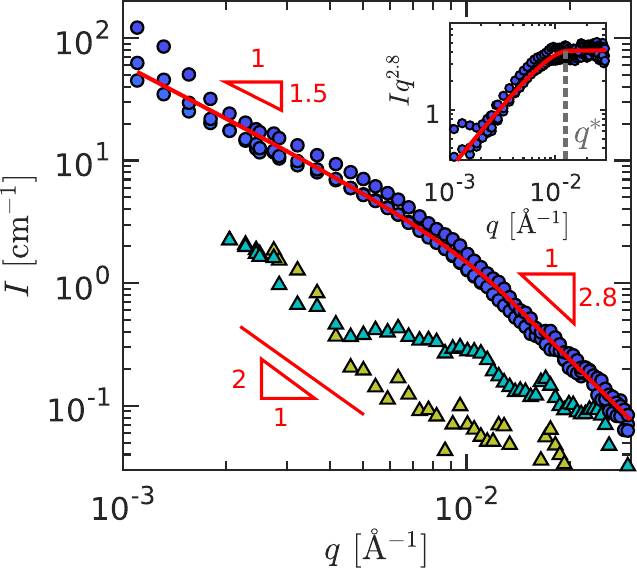}
    \caption{Small angle neutron scattering intensity $I$ as a function of the momentum transfer $q$ for samples prepared at $\mathrm{pH}<\mathrm{pH_c}$ ($\circ$) and $\mathrm{pH}>\mathrm{pH_c}$ ($\triangle$). Same color code as in Fig.~\ref{fig:transition_sol_gel}. The red solid line is the best fit by a Guinier-Porod model\cite{hammouda2010new}, see SI. Inset: Kratky-like representation $I q^{2.8}$ vs.~$q$ of the scattering data of the gel phase presented in the main graph. The vertical dashed line highlights $q^*$.}
    \label{fig:SANS}
\end{figure}

\begin{figure*}[t]
    \includegraphics[width = 0.8\textwidth]{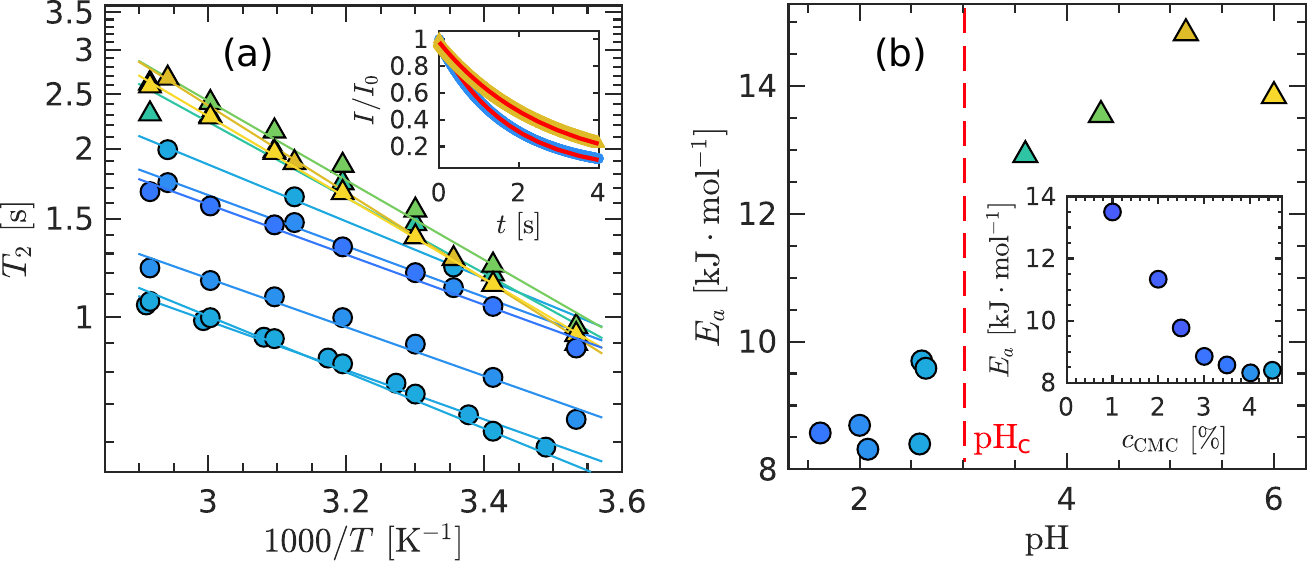}
    \caption{(a) Transverse NMR relaxation time $T_2$ of CMC solutions at $c_{\rm CMC} = 3 \%$ and different pH as a function of the inverse of the temperature plotted in semilogarithmic scale. Solid lines correspond to the best Arrhenius fit of the data $T_2 \sim \exp (-E_a  / R T) $. Same color code as in Fig.~\ref{fig:transition_sol_gel}. Inset: intensity $I$ of the transverse magnetic field as a function of time $t$ (normalized by its initial value $I_0$). Solid lines correspond to the best fit of the data with an exponential fit $I / I_0 = \exp(-t/T_2)$. (b) Parameter $E_a$ obtained from the Arrhenius fit in (a) as a function of pH at fixed $c_{\rm CMC} = 3 \%$. Inset: $E_a$ vs.~$c_{\rm CMC}$ at fixed $c_{\rm HCl} = 0.15 \rm M$. Triangles (resp.~disks) correspond to the liquid (resp.~gel) phase. Same color code as in Fig.~\ref{fig:transition_sol_gel}.}
    \label{fig:RMN}
\end{figure*}

To further elucidate the self-assembly of CMC upon acidification, we use small-angle neutron scattering (SANS) and report the scattering intensity spectra $I(q)$ in Fig.~\ref{fig:SANS}. For viscoelastic liquids, i.e., at $\rm pH >3$, the signal-to-noise ratio is low, and $I(q)$ decays roughly as $q^{-2}$. This high-$q$ (Porod) regime is reminiscent of non-interacting Gaussian polymer chains in a solvent. For $\rm pH <3$, i.e., upon gelation, the scattered intensity increases dramatically, showing that CMC gels scatter neutrons much more efficiently than their liquid counterparts. Moreover, two important observations can be made. First, the Porod regime at low pH switches to $I\sim q^{-2.8}$ indicating the presence of relatively dense objects together with a gradient of scattering length density at the interface with the background \cite{beaucage1994structural}. More precisely, the exponent $-2.8$ represents an intermediate case between Gaussian chains ($I\sim q^{-2}$), where the exponent is related to the volume fractal dimension of the chain, and smooth particles ($I\sim q^{-4}$), where the exponent is related to the surface fractal dimension $D_s$ according to $I\sim q^{-(6-D_s)}$ (where $D_s=2$ for smooth particles). This result is compatible with the presence of fringed micelles\cite{Liebert:2001} that are characterized by a dense core formed by the aggregation of hydrophobic CMC strands surrounded by a sparse, hairy shell of polymer chains. Second, fitting the scattering intensity of gel samples with an empirical Guinier-Porod model \cite{hammouda2010new} enables us to extract the inflection point of the scattering intensity, providing $q^* \simeq 8.4\ 10^{-3} \AA^{-1}$ (see also inset in Fig.~\ref{fig:SANS}) corresponding to a length scale $d^*=2\pi/q^*\simeq 75$~nm that we assign to the diameter of the fringed micelles section. At $q<q^*$, the scattering intensity scales as $I\sim q^{-1.5}$ suggesting that the micelles can be seen as flexible rod-like objects characterized by an aspect ratio larger than about 10, in good agreement with the schematic view proposed in Ref.~\onlinecite{Lopez:2018} for CMC of low DS, spontaneously gelling in concentrated solutions. Here, the largest characteristic size of the fringed micelles is larger than 600~nm since no plateau is visible at the smallest accessible $q$ values.
Focusing on gel samples in the low $q$ range, the scattering intensity increases slightly but monotonically with decreasing pH, which suggests that the mass of the fringed micelles, or the fractal dimension of the network they form, tend to grow progressively even well beyond the gel point. This result indicates that while most of the hydrophobic patches tend to aggregate at the sol-gel transition, forming the so-called fringed micelles, a fraction of them remain available in the gel state.

Finally, we examine the sol-gel transition from the \textit{solvent perspective} using molecular probes.
Low-field NMR experiments allow us to assess the dynamics of the water molecules\cite{note1} for different pH. We observe that irrespective of the sample pH, the spin relaxation following a 90$^\circ$ pulse (see \textit{Experimental} section) is a decreasing function of time [see inset in Fig.~\ref{fig:RMN}(a)]. The data are fitted with a single mode decreasing exponential function, i.e., $I(t)=I_0\exp(-t/T_2)$, where $I_0$ is the signal intensity at $t=0$. In contrast to similar measurements on colloidal dispersions,\cite{le1998quantitative} or hydrogels based on methylcellulose \cite{besghini2023time}, chitosan \cite{capitani2001water}, biological tissues \cite{abrami2018use}, and synthetic homo- and co-polymers \cite{ghi2002h,abrami2014physical}, we find that an excellent fit of the relaxation data can be achieved with a single exponential term. CMC samples in both the sol and gel phase are therefore characterized by a \textit{single} transverse (spin-spin) relaxation time $T_2$.
Repeating the same experiments at various temperatures reveals that $T_2$ increases for increasing temperature [see Fig.~\ref{fig:RMN}(a)], which we attribute to a higher mobility of the water molecules regardless of their environment. More specifically, $T_2 (T)$ is well fitted by an Arrhenius law, i.e.,  $T_2=T_2^0\exp(-E_a/RT)$, yielding an apparent activation energy $E_a$ for each sample composition. Remarkably, $E_a$ shows an abrupt jump at the sol-gel transition, from $E_a \simeq 14~\rm kJ.mol^{-1}$ for liquid-like samples ($\mathrm{pH}>3$) to $E_a \simeq 9~\rm kJ.mol^{-1}$ for solid-like samples ($\mathrm{pH}<3$).  
Moreover, as shown in the inset of Fig.~\ref{fig:RMN}(b), a decrease in the CMC concentration at fixed $\mathrm{pH}<\mathrm{pH}_c$ results in a continuous increase of $E_a$ up to about $14~\rm kJ.mol^{-1}$, similar to the value measured for $\mathrm{pH}>\mathrm{pH}_c$ at $c_{\rm CMC}=3\%$.
Note, however, that the apparent activation energy $E_a$ cannot be understood as a bonding energy as often proposed to rationalize Arrhenius dependence observed in rheology.\cite{leibler1991dynamics, baeza2016network, nian2023dynamics} Here, low $T_2$ values accompanied by low $E_a$ are indicative of the formation of rigid systems, in which the dynamics of the water molecules are slowed down and poorly sensitive to temperature. Such a picture is in good qualitative agreement with the formation of a percolated network at low pH, which causes a decrease \textit{in bulk} of the average water molecules diffusivity. On the other hand, increasing the pH or decreasing the CMC concentration results in looser networks of fringed micelles, and eventually non-percolated systems, where the mobility of water molecules is enhanced and effectively more sensitive to temperature. 

Put together, our observations offer a comprehensive picture of the acid-induced sol-gel transition. 
Upon acidification, CMC assemble into fringed micelles that serve as precursors for the formation of a percolated network made of CMC fibers, as evidenced by SANS measurements and strongly reminiscent of the fibrous microstructure reported for heat-induced methylcellulose gels.\cite{Lott:2013,besghini2023time} The sol-gel transition, which relies on the binding of hydrophobic patches spread along the CMC molecules, is \textit{abrupt} both from the perspective of the gel microstructure (SANS experiments) and from that of the solvent (low-field NMR experiments). These observations support the picture of a sol-gel transition taking place at a critical density of activated hydrophobic patches, as introduced above to account for the equation of the boundary separating the sol from the gel region in the phase diagram of Fig.~\ref{fig:transition_sol_gel}(b). Beyond the gel point, decreasing the pH increases the number of hydrophobic patches involved in the gel microstructure, which also induces a change in the CMC conformation. These additional crosslinkers contribute to reinforcing the gel elastic properties over a limited range of pH beyond which the elastic modulus $G_0$ saturates towards a plateau [see, e.g.,  Fig.~\ref{fig:Rheo_gel}(a) at pH~$<1.5$ for a 3\% CMC solution]. This suggests that some hydrophobic patches, although active, do not contribute to the gel elastic properties, either due to steric constraints or to intramolecular binding. Our results provide solid ground for designing CMC-based soft composites with more complex compositions.

\section{Experimental}
\label{sec:Exp}

\subsubsection{Sample preparation}

Samples are prepared by dispersing sodium carboxymethyl cellulose (Sigma Aldrich, $M_w=250$~kg.mol$^{-1}$ and DS$~=0.9$) in deionized water. {DS was independently confirmed by high-field NMR spectroscopy (see SI), while $M_w$ was determined to be $242.2~\rm kg.mol^{-1}$ in Ref.~\onlinecite{Komorowska:2017}}. Stock solutions up to 5\% wt.~are prepared and stirred at room temperature for 48~h until homogeneous, before diluting them with a hydrochloric acid solution (Sigma Aldrich). Samples are then placed on a bottle roller for 7~days before being tested {at 10 days to limit the impact of ageing (see SI)}. Final CMC solutions concentrations span from 1~\% to 4.5~\%. We used a 1~M HCl solution for all the dilutions except for the most concentrated CMC solutions (i.e., 4 and 4.5~\%) for which we used a 12~M HCl solution. We checked there is no impact of the HCl solution concentration for a sample at $c_{\rm CMC} = 1\%$ and $\mathrm{pH}=1$. The pH of the samples is measured with a pH-meter (Mettler Toledo SevenCompact) and spans between 0 and 7. 

\subsubsection{Rheometry}
The rheological properties of NaCMC aqueous solutions are determined with a cone-and-plate geometry (angle 2$^\circ$, radius 20~mm and truncation 46~$\mu$m) connected to a strain-controlled rheometer (ARES-G2, TA Instruments). The cone and plate are both sandblasted and display a surface roughness of about 1~$\mu$m to prevent wall slip. Samples are loaded in the shear cell and maintained at constant temperature $T=22^\circ$C during the duration of the test, thanks to a Peltier modulus placed under the bottom plate. The rheological protocol applied to all the samples is divided into three consecutive steps: ($i$) a preshear at $\dot \gamma=50$~s$^{-1}$ for 3~min to erase the loading history, and rejuvenate the sample; ($ii$) a 20~min recovery during which we monitor the sample linear viscoelastic properties by applying small amplitude oscillations $\gamma=1$\% at 1~Hz, {and which allows us to conclude that a steady state is reached typically beyond 100~s}; ($iii$) a frequency sweep performed at $\gamma=1$\% to determine the linear viscoelastic spectrum of the sample over four decades in frequency.

\subsubsection{Small-angle neutron scattering}
Small-angle neutron scattering (SANS) experiments were performed on the beamline D22 at Institut Laue Langevin (Grenoble, France). Samples were loaded into 1~mm thick Hellma cells made of quartz and measured at room temperature. Two configurations were used to maximize the accessible $q$ range. The wavelength was set either to $\lambda=6$~\AA~or $\lambda=11.5$~\AA, and the sample-detector distance was fixed {at 17.6~m for the back detector and 1.4~m for the front detector}. Samples were systematically measured in heavy water ($\rm D_2O$) to optimize the contrast between CMC and its background. The scattering length density of CMC is ca.~$\rho_{\rm CMC}=1.0\times 10^{10}\ \rm cm^{-2}$ (assuming a density of $1.6\ \rm g.cm^{-3}$) while that of $\rm D_2O$ is ca.~$\rho_{\rm D_2O}=6.4\times 10^{10}\ \rm cm^{-2}$.  Data were appropriately corrected by the transmission, {empty cell, background,} and thickness to obtain absolute units ($\rm cm^{-1}$)  {thanks to a direct beam measurement}.\\

\subsubsection{Time-domain $^1$H NMR Measurements}
Measurements of $T_2$ transverse relaxation time were performed on CMC solutions using a low-field NMR pulsed spectrometer (Bruker Minispec mq20), at a 20 MHz proton resonance frequency.  Samples were placed into tube with a diameter of 1~cm, and filled up to a height of ca.~3~mm, to avoid spatial heterogeneities of the magnetic field. The temperature was increased from 283~K up to 343~K with a BVT 3000 heater working with nitrogen gas. Prior to each experiment, the temperature was stabilized for 10~min. The measurements followed a Car-Purcell-Meiboom-Gill (CPMG) echo train procedure enabling access to the long transverse relaxation time of the protons present in water (the signal coming from the polymer protons was neither measurable nor investigated). The echo time was set to 1~ms, and the signal at short time was systematically adjusted to be between 60\% and 100\% intensity, ensuring optimal statistics without any risk of saturation.\\

\begin{acknowledgement}
The authors thank L.~Morlet-Decarnin and S.~Denis-Quanquin for preliminary tests on low-field NMR, and J.~Bauland, E. Del Gado, E.~Freyssingeas, F.~Li\'enard, and P.D. Olmsted for fruitful discussions, {as well as C.G.~Lopez for pointing out relevant references}. This work was supported by the LABEX iMUST of the University of Lyon (ANR-10-LABX-0064), created within the ``Plan France 2030" set up by the French government and managed by the French National Research Agency (ANR). All the authors acknowledge the Laue Langevin Institute (ILL) for the provision of the neutron beam facilities and the assistance in using {beamline D22}. Corresponding data are available at the following 10.5291/ILL-DATA.9-12-702. This work benefited from the use of the SasView application, originally developed under NSF award DMR-0520547. SasView contains code developed with funding from the European Union’s Horizon 2020 research and innovation programme under the SINE2020 project, grant agreement No 654000.
\end{acknowledgement}

\begin{suppinfo}

Supporting information includes ($i$) the determination of DS by high-field NMR spectroscopy, ($ii$) the rheological properties of chemically degraded samples,  ($iii$) a characterization of the aging dynamics of gel samples,  ($iv$) a model for the sol-gel boundary in Fig.~\ref{fig:transition_sol_gel} in the main text, ($v$) the individual linear viscoelastic spectra of 3\% CMC gels at various pH, and ($vi$) the expression of the fit function for the SANS intensity measured in the gel phase and reported in Fig.~\ref{fig:SANS} in the main text.

\end{suppinfo}

\clearpage

\onecolumn
\setcounter{equation}{0}
\setcounter{figure}{0}
\global\def\thefigure{S\arabic{figure}}
\setcounter{table}{0}
\global\def\thetable{S\arabic{table}}

\begin{center}
\LARGE{Supplementary Information}
\end{center}

\section{Determination of DS by high-field NMR spectroscopy}

{To confirm the value of the degree of substitution (DS) for our sample (a priori, $\mathrm{DS}=0.9$ based on the information provided by Sigma Aldrich), we have performed NMR experiments on a 6\% CMC sample prepared in D$_2$O. The result, illustrated in Fig.~\ref{fig:RMN}, yields a value of $\mathrm{DS}=0.88$ with an uncertainty estimated to be about $3\%$. This value is in excellent agreement with that provided by Sigma Aldrich. 
}

\begin{figure*}[h!]
    \centering
    \includegraphics[width = 0.9\textwidth]{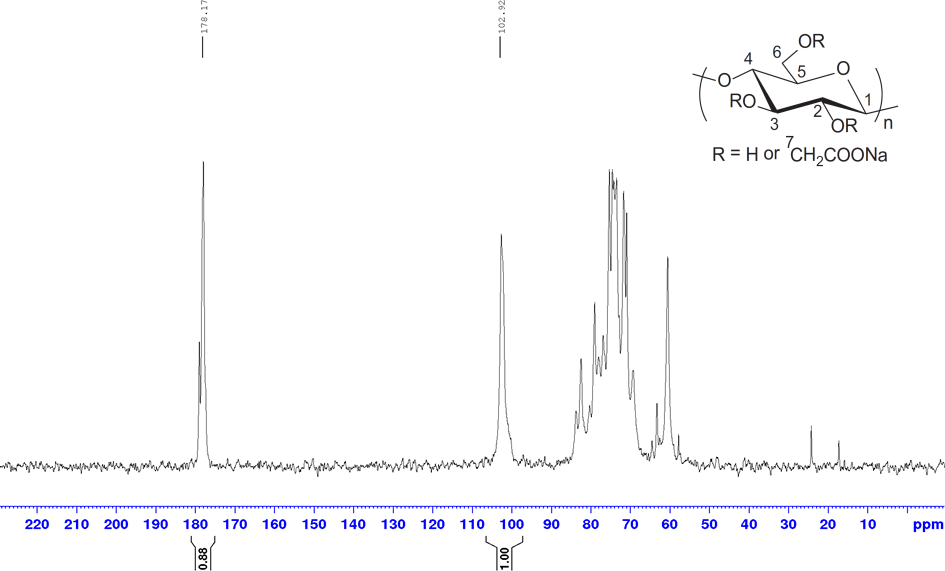}
    \caption{{Spectrum $^{13}$C-NMR inverse gate decoupling, with a pulse angle 70$^{\circ}$, recycling delay 10~s, 5120~scans, $T=343$~K of a 6\% CMC sample prepared in D$_2$O. The ratio between integrals of the CO signal (178 ppm) and the C1 signal (102 ppm) provides a direct estimate of the degree of substitution. The measurement was performed with an Avance II Ascend 400 MHz (Bruker, USA).}}
    \label{fig:RMN}
\end{figure*}

\clearpage

\section{Rheological properties of chemically degraded samples}

\begin{figure*}[!h]
    \centering
    \includegraphics[width = 0.8\textwidth]{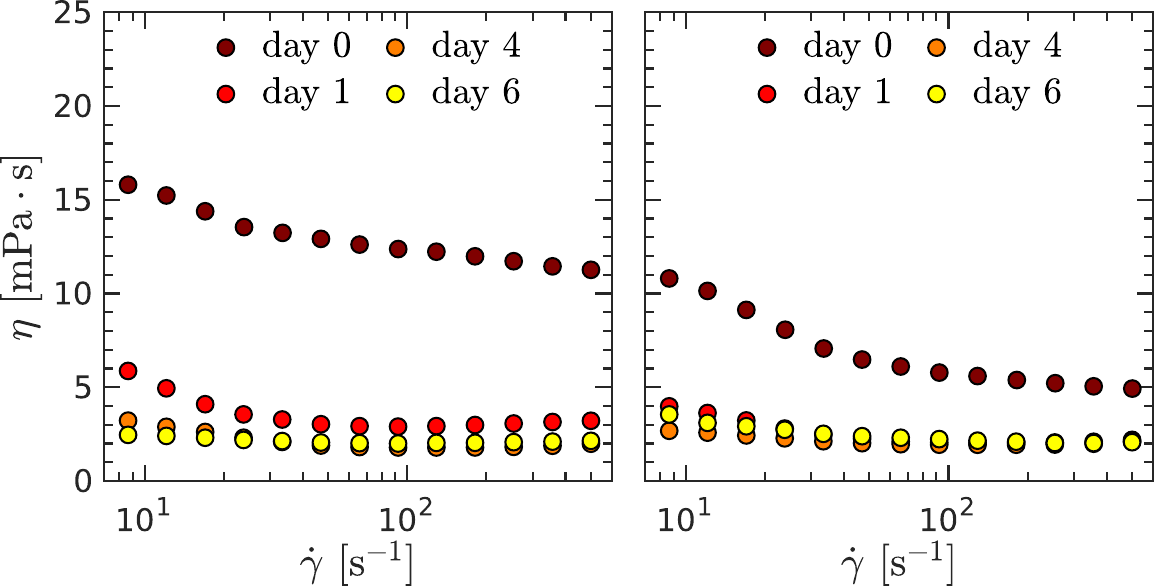}
    \caption{{Viscosity $\eta$ vs shear rate $\dot \gamma$ of two CMC solutions (a) $c_{\rm CMC} = 1.9\%$ and pH~$=0$, (b) $c_{\rm CMC} = 1.4\%$ and pH~$=0$. Colors encode the sample age, i.e., the time elapsed since the sample was prepared. Rheological measurements were performed in a cone-and-plate geometry, angle 2$^\circ$, diameter 50~mm, mounted on a stress-controlled rheometer, MCR 302 (Anton Paar). }}
    \label{fig:netwonien}
\end{figure*}

{The degraded CMC solutions are translucent, and their viscosity drops by several orders of magnitude compared to gels and non-degraded viscoelastic liquid-like samples. The viscosity of the degraded samples does not depend on the applied shear rate, i.e., these samples display a Newtonian behaviour. We used Newtonian behaviour as a criterion to classify these samples as degraded. While degraded samples do not show any time-dependent rheological properties such as thixotropy, they display \textit{irreversible aging}, i.e., their viscosity decreases with increasing sample age on a typical timescale of one or two days following their preparation, as illustrated in Fig.~\ref{fig:netwonien}. We attribute this aging to the slow chemical process during which polymer chains are broken down into smaller polymers.\cite{Cheng:1999}
Therefore, we conclude that visual inspection, together with viscosity measurements, provide robust qualitative and quantitative criteria for identifying the degraded samples, whose rheology is Newtonian and comparable to that of water after a couple of days only. }\\

\clearpage

\section{Aging dynamics of gel samples}

\begin{figure*}[!h]
    \centering
    \includegraphics[width = 0.45\textwidth]{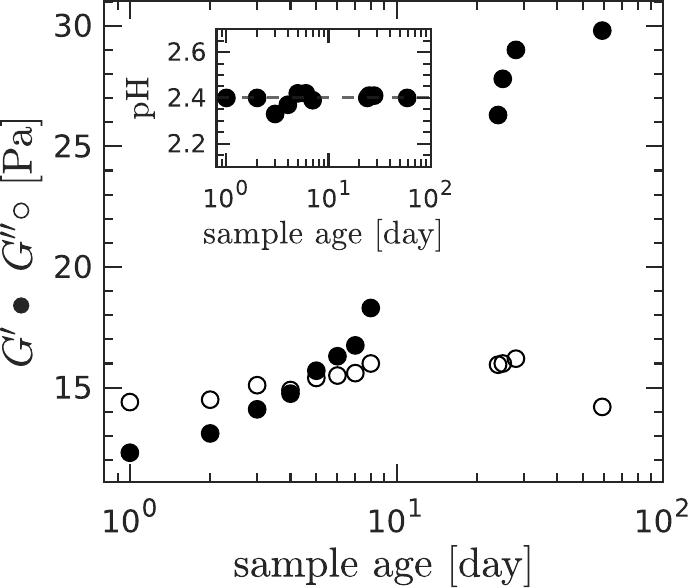}
    \caption{{Elastic modulus $G'$ and viscous modulus $G''$ as a function of the sample age, i.e., of the time elapsed since its preparation, for a prototypical gel sample ($c_{\rm CMC} = 3\%$ and pH~$=2.4$). In practice, each point corresponds to a sample taken from a batch kept at room temperature and placed in a cone-and-plate geometry (50~mm diameter, angle 2$^\circ$, and truncation 211~$\mu$m). Measurement performed at $\gamma_0=1\%$ and $f=1$~Hz. The elastic properties strengthen as a function of the sample age and reach a steady state beyond 20 days. Inset: pH of the sample as a function of the sample age. The pH is insensitive to sample age.}}
    \label{fig:aging}
\end{figure*}

{We tested the time dependence of the rheological properties of CMC samples in the gel phase. The linear viscoelastic properties of a prototypical gel, i.e., a 3\% CMC sample at pH=2.4 are reported in Figure~\ref{fig:aging} over the course of 2 months. We observe that the sample elastic modulus strengthens in a logarithmic fashion as a function of the sample age, increasing from 10~Pa to 30~Pa within 20 days, while $G''$ remains roughly constant, about 15~Pa. This observation, reminiscent of previous reports on HCMC gels,\cite{Hakert:1989} confirms that the rheological properties of the gel samples are moderately time-dependent. All the viscoelastic properties used to build the phase diagram reported in Figure~1(b) in the main text were measured at least 10 days after sample preparation. Figure~\ref{fig:aging} confirms that neither our conclusions nor the phase diagram are affected by sample aging. In particular, we emphasize that we have not observed any sample from the gel phase, whose rheological properties would slowly degrade as a function of time. }\\

\clearpage

\section{Model for the sol-gel boundary}

Here, we explicit the assumptions of the model used to describe the sol-gel boundary of the phase diagram and reported as a red curve in Fig.~1 in the main text. This model constitutes a simple description of the sol-gel transition, based on  experimental elements available, and on the following assumptions:\\

\textbf{$\bullet$ First assumption.} The sol-gel transition occurs when the total concentration of hydrophobic patches, i.e., the number $n$ of patches per unit volume, exceeds a given threshold $n_{\rm crit}$. As the CMC solution is in the semi-dilute entangled regime, we assume that $n_{\rm crit}$, which corresponds to a given average functionality of the patches, neither depends on the CMC content nor on the pH.  
In other words, we neglect the changes in the interaction energies associated with a change in polymer or HCl concentrations. 
Moreover, we define $f_{\rm OFF}$ (= $1  -f_{\rm ON}$) as the probability for a hydrophobic patch to be disconnected from the elastically active network (e.g., patches connected within the same polymer or patches that are isolated). This fraction is dictated by the conformation of the polymer and, in particular, by its degree of coiling. Therefore,  $f_{\rm OFF}$ should mainly depend on the pH, and presumably much less on the CMC concentration $c_{\rm CMC}$.
Finally, the hydrophobic crosslinks have a finite lifetime, i.e., on average, there is a fraction $f_{\tau}$ of patches that are effectively connected at a given time. 
In that framework, the sol-gel transition occurs when 
\begin{align}
     f_{\rm ON} f_{\tau } n = n_{\rm crit}\,. \label{eq:n_transition}
\end{align}

\textbf{$\bullet$ Second assumption.} For the sake of simplicity, we consider that the probability for a hydrophobic patch to be active is constant, i.e., $f_{\rm ON} = cst $. This assumption is supported by the fact that (i)~the weight fraction of CMC in the gels does not exceed 4 wt.\%, so that the system remains in the semi-dilute regime, and (ii)~we focus on the vicinity of the sol-gel transition so that the coiling of the polymer should be quite similar regardless of the CMC concentration or the pH. Furthermore, we consider that the active hydrophobic patches are active most of the time, i.e., $f_{\tau }\simeq1$. The typical values for $f_{\rm ON}$ and $f_{\tau }$ are actually not important, as they are considered constant, which is equivalent to consider an effective $n_{\rm crit}^{\rm eff} = n_{\rm crit} / (f_{\rm ON} f_{\tau })$, and set $f_{\rm ON} = f_{\tau }=1$ thereafter. For clarity, we shall still keep $f_{\rm ON}$ and $f_{\tau }$ explicitly below.

\bigskip 
\textbf{$\bullet$ Third assumption.} We now express the concentration of hydrophobic patches $n$ as a function of $c_{\rm CMC}$ and pH to compute the equation of the sol-gel boundary. We propose the following form, introducing $c_{\rm COOH}$ the concentration of substituted groups that are in their acidic form:
\begin{align}
     n = A c_{\rm CMC} + B c_{\rm COOH} \,,
\end{align}
where $A$ and $B$ are positive prefactors. Both concentrations $c_{\rm CMC}$ and $ c_{\rm COOH}$ are expressed in mol.L$^{-1}$ (proportional to the number of entities per unit volume). The prefactor $A$ corresponds to the number of patches introduced by the CMC itself without altering the pH of the solution; $A$ depends mostly on the degree of substitution (DS) of the CMC, and on the molecular weight $M_w$ of the CMC. In contrast, $B$ denotes the additional number of hydrophobic patches when a COO$^{-}$ group is protonated. Hence, $B$ should depend not only on DS and $M_w$, but also on the patch density along the polymer chain, if many COOH groups are formed in a dense environment of patches, they might increase the length of already existing patches rather than creating a new one. \\

\textbf{$\bullet$ Fourth assumption.} We  consider $A$ and $B$ to be constant, independent of $c_{\rm CMC}$ and pH. $ c_{\rm COOH} $ is linked to the pH through the following two equations that capture respectively the atom conservation (when a mole of CMC is introduced, it introduces $ \xi = N \times DS $ moles of COOH or COO$^{-}$, with $N$ the number of monomers, proportional to $M_w$), and the acido-basic chemistry with an acidity constant $K_A$:
\begin{align}
    c_{\rm COOH} + c_{\rm COO^{-}}  = \xi \ c_{\rm CMC}\,, \\
    K_A = \dfrac{10^{-\rm pH} \ c_{\rm COO^{-}}}{ c_{\rm COOH} }\,.
\end{align}

Combined with Eq. (2), this leads to the following expressions of $c_{C\rm OOH}$ and $n$:
\begin{align}
    c_{\rm COOH} = \dfrac{\xi \ c_{\rm CMC}}{1 + 10^{\rm pH - pK_A}} \label{eq:cooh}\,, \\
    n = c_{\rm CMC} \left(A +  \dfrac{B \xi}{1 + 10^{\rm pH - pK_A}} \right). \label{eq:n_dependence}
\end{align}

Combining Eqs.~\eqref{eq:n_transition} and \eqref{eq:n_dependence} then yields the equation for the sol-gel boundary:
\begin{align}
   \mathrm{pH_c} = \mathrm{pK_A} + \log_{10} \dfrac{ (A + B \xi ) f_{\rm ON} f_\tau c_{\rm CMC} - n_{\rm crit} }{n_{\rm crit} - A f_{\rm ON} f_\tau c_{\rm CMC} }\,.
\end{align}

\textbf{$\bullet$ Fifth assumption.} $n_{\rm crit}$ is related to the critical CMC concentration below which no gel can be obtained, irrespective of the pH. Here, we observe experimentally that  $c_{\rm crit} \simeq 1~\%$. In Eq.~(\ref{eq:cooh}), the largest possible value of $c_{\rm COOH}$ is $\xi c_{\rm CMC}$, for which $n \simeq c_{\rm CMC}  \left( A + B \xi \right) $, so that an estimation of $n_{\rm crit}$ can be expressed as follows:
\begin{align}
    n_{\rm crit} \simeq f_{\rm ON} f_{\tau } \left( A + B \xi \right) c_{\rm crit}
\end{align}
Based on this estimate, we can express the equation of the sol-gel boundary as follows:
\begin{align}
    \mathrm{pH_c} &= \mathrm{pK_A} + \log \dfrac{ c_{\rm CMC} - c_{\rm crit}}{ c_{\rm crit} - \dfrac{A}{A + B \xi }\, c_{\rm CMC} } \\
    &= \mathrm{pK_A} + \log \dfrac{ c_{\rm CMC} - c_{\rm crit}}{ c_{\rm crit} - \alpha c_{\rm CMC} }\,. \label{eq:finale}
\end{align}
with $\alpha = {A}/{(A + B \xi)} < 1$.

Finally, we use Eq.~\eqref{eq:finale} to fit the boundary of the sol-gel phase diagram reported in Fig.~1 in the main text. The best agreement with the experimental data (reported in red in Fig.~1) is obtained for $\alpha=0.18$ and ${\rm pK_{A}}=2.5$, consistently with ${\rm pK_{A}}$ values reported in  Ref.~\cite{Hoogendam:1998}

\bigskip 

\textbf{$\bullet$ Final remark.} Note that in Eq.~\eqref{eq:finale} $\rm pH_c$ diverges when $c_{\rm CMC} = c_{\rm crit} / \alpha$, which means that above this concentration, a gel will form regardless of the pH, which is indeed what is observed in the literature for similar DS and $M_w$ of the CMC\cite{Lopez:2021,Lim:2015}.


\clearpage

\section{Linear viscoelastic spectra of a 3\% CMC gel at various pH}

\begin{figure*}[!h]
    \centering
    \includegraphics[width = 1\textwidth]{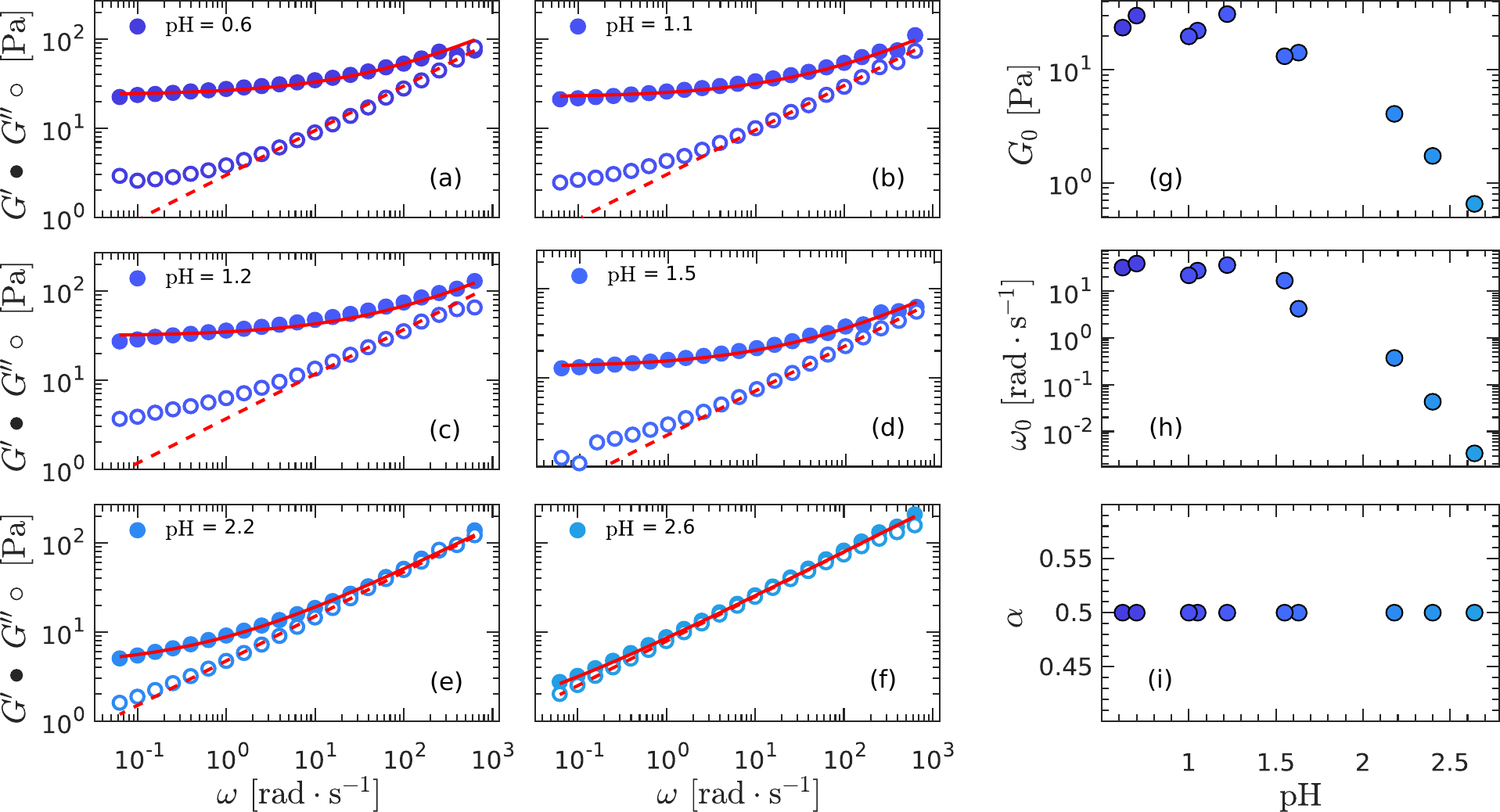}
    \caption{{(a-f) Viscoelastic spectra of CMC solutions at fixed polymer content $c_{\rm CMC} = 3\%$ and pH ranging from 0.6 to 2.6. The solid  (resp. dashed) red curve corresponds to $G'$ (resp. $G''$) when fitting the complex modulus to eq.~(1) in the main text. The three fitting parameters are plotted against the pH: (g)~$G_0$, (h)~$\omega_0$, and (i)~$\alpha$. The latter is fixed to $\alpha = 0.5$, as explained in the main text.}}
    \label{fig:KVF}
\end{figure*}

Figure~\ref{fig:KVF} displays the individual viscoelastic spectra, i.e., $G'$ and $G''$ vs.~$\omega$ for the ten samples of composition $c_{\rm CMC} = 3~\%$, and pH varying between 0.6 and 2.6 [Fig.~\ref{fig:KVF}(a)-(f)]. The graphs also display the best fits of the data by the fractional Kelvin-Voigt model introduced in the main text, as well as the evolution of the fitting parameters $G_0$, $\omega_0$, and $\alpha$ (the latter being set constant equal to 0.5) as a function of the pH [Fig.~\ref{fig:KVF}(g)-(i)].\\

\clearpage

\section{Fit function for the SANS intensity in the gel phase}
 The empirical Guinier-Porod fit function used in Fig.~3 in the main text was developed by B.~Hammouda,\cite{hammouda2010new} and more recently added to the SasView application\cite{sas}. It is defined as follows:
 \begin{equation}
    \begin{array}{lcc}
    I(q<q^*)={G}/{q^s}\exp\left(\frac{-q^2R_g^2}{3-s}\right),\\
    I(q>q^*)={D}/{q^m}\,.
    \end{array}
\end{equation}
where $s$ and $m$ are the power-law exponents at low $q$ (Guinier regime) and high $q$ (Porod regime), respectively. $R_g$ is the apparent radius of gyration that relates to $q^*$, $m$, and $s$ through:
 \begin{equation}
    q^*=\frac{1}{R_g}\sqrt{(m-s)(3-s)/2}\,.
\end{equation}
\noindent Moroever, to ensure continuity of $I(q)$, the prefactors $G$ and $D$ in the low-$q$ and high-$q$, respectively, are related to each other through: 
 \begin{equation}
    D=G\ \exp\left(\frac{-q^{*2}R_g^2}{3-s}\right)q^{*m-s}=
    \frac{G}{R_g^{m-s}}\ \exp\left(-\frac{m-s}{2}\right) \left(\frac{(m-s)(3-s)}{2}\right)^{\frac{m-s}{2}}.
\end{equation}


\begin{mcitethebibliography}{48}
\bibitem[not()]{note1}
In fact, the recorded magnetization is directly proportional to the number of
  protons in the sample, the signals measured on the hydrogels are thus
  strongly dominated by the water molecules.\relax
\mciteBstWouldAddEndPunctfalse
\mciteSetBstMidEndSepPunct{\mcitedefaultmidpunct}
{}{\mcitedefaultseppunct}\relax
\EndOfBibitem{}
\end{mcitethebibliography}

\providecommand{\latin}[1]{#1}
\makeatletter
\providecommand{\doi}
  {\begingroup\let\do\@makeother\dospecials
  \catcode`\{=1 \catcode`\}=2 \doi@aux}
\providecommand{\doi@aux}[1]{\endgroup\texttt{#1}}
\makeatother
\providecommand*\mcitethebibliography{\thebibliography}
\csname @ifundefined\endcsname{endmcitethebibliography}
  {\let\endmcitethebibliography\endthebibliography}{}

\end{document}